\documentstyle[12pt]{article}

\def\ben{\begin{enumerate}}  \def\een{\end{enumerate}}
\def\beq{\begin{equation}}   \def\eeq{\end{equation}}
\def\bea{\begin{eqnarray}}  \def\eea{\end{eqnarray}}
\def\nn{\nonumber}
\def\noi{\noindent}

\def\lsim{\raise0.3ex\hbox{$<$\kern-0.75em\raise-1.1ex\hbox{$\sim$}}}
\def\gsim{\raise0.3ex\hbox{$>$\kern-0.75em\raise-1.1ex\hbox{$\sim$}}}

\begin{document}

\begin{center}
{\large \bf Uraltsev Sum Rule in Bakamjian-Thomas Quark Models}
\par

\vspace{1 truecm}

{\bf A. Le Yaouanc, L. Oliver, O. P\`ene and J.-C. Raynal}\\
{\it Laboratoire de Physique Th\'eorique}\footnote{Unit\'e Mixte de Recherche
UMR 8627 - CNRS }\\    {\it Universit\'e de Paris XI, B\^atiment 210, 91405
Orsay Cedex, France}
\par \vskip 5 truemm
{\bf V. Mor\'enas} \\
{\it Laboratoire de Physique Corpusculaire}\\
{\it Universit\'e Blaise Pascal - CNRS/IN2P3, F-63000 Aubi\`ere Cedex, France}
   \end{center}

\vspace{0.5 truecm}
\begin{abstract}
We show that the sum rule recently proved by Uraltsev in the heavy 
quark limit of QCD holds in relativistic
quark models \`a la Bakamjian and Thomas, that were already shown to 
satisfy Isgur-Wise scaling and Bjorken sum
rule. This new sum rule provides a {\it rationale} for the lower 
bound of the slope of
the elastic IW function $\rho^2  \geq  {3 \over 4}$ obtained within 
the BT formalism
some years ago. Uraltsev sum rule suggests an inequality 
$|\tau_{3/2}(1)|  > |\tau_{1/2}(1)|$. This difference is
interpreted in the BT formalism as due to the Wigner rotation of the 
light quark spin, independently of a possible LS
force. In BT models, the sum rule convergence is very fast, the $n = 
0$ state giving the essential contribution in most
of the phenomenological potential models. We underline that there is 
a serious problem, in the heavy quark limit of QCD,
between theory and experiment for the decays $B \to D^*_{0,1}(broad) 
\ell \nu$, independently of any model calculation.
   \end{abstract}

\vspace{1 truecm}

\noi LPT Orsay 01-20 \par
\noi March 2001
 
\newpage
\pagestyle{plain}

\section{Introduction.}
\hspace*{\parindent}
	Recently, N. Uraltsev \cite{1r} has established, in the heavy 
quark limit of QCD, a new
sum rule. The demonstration of the sum rule (SR) follows from the OPE 
applied to the
scattering amplitude $T(\varepsilon,{\bf v},{\bf v} - {\bf v}')$ in 
the Shifman-Voloshin
limit. The function $T(\varepsilon,{\bf v},{\bf v} - {\bf v}')$, 
where $\varepsilon$ is
the energy variable ($\varepsilon = 0$ for elastic transitions of a 
free quark), is the
Fourier transform of the expectation value

\beq
	< B^*({\bf v} - {\bf v}')|T(J^+(0)J(x))|B^*(0) >
\label{1e}
\eeq

\noi where the initial state is at rest and the final state has a momentum
$m_Q({\bf v} - {\bf v}')$, $-m_Q{\bf v}$ being the momentum transfer 
carried by the
intermediate states. The novelty in Uraltsev procedure is to allow a 
momentum for the
final state in (\ref{1e}). Then, the function $T(\varepsilon,{\bf 
v},{\bf v} - {\bf
v}')$ can be decomposed into symmetric and antisymmetric parts 
$h_{\pm}(\varepsilon)$
in ${\bf v}$, ${\bf v}'$. The zero order moment of $h_+(\varepsilon)$ leads to
Bjorken SR \cite{2r} involving $\rho^2$, the slope of the elastic IW 
function $\xi (w)$~:

\beq
\label{2e}
		\rho^2 = {1 \over 4}   +  \sum_n 
|\tau^{(n)}_{1/2}(1) |^2 + 2 \sum_n
|\tau^{(n)}_{3/2}(1) |^2 			 \eeq

\noi while the zero order moment of $h_-(\varepsilon)$ leads to the 
new SR \cite{1r}~:

  \beq
\label{3e}
		\sum_n  |\tau^{(n)}_{3/2}(1) |^2 - \sum_n 
|\tau^{(n)}_{1/2}(1) |^2 = {1 \over 4}
\quad . \eeq

\noi From (2) and (3) one gets the lower bound

\beq
\label{4e}
					\rho^2  \geq  {3 \over 4} \quad .
\eeq

	The simple relations that come out immediately from (\ref{2e}) and (\ref{3e}),
\bea
\label{5e}
&&		\sum_n  |\tau^{(n)}_{3/2}(1) |^2 = {\rho^2 \over 3} \\
&&		\sum_n  |\tau^{(n)}_{1/2}(1) |^2 = {1 \over 3} \left 
( \rho^2 - {3 \over 4}
\right ) 				 \label{6e}
\eea

\noi deserve a comment. One can see that $\sum\limits_n 
|\tau^{(n)}_{3/2}(1) |^2$ is
proportional to $\rho^2$ and that $\sum\limits_n  |\tau^{(n)}_{1/2}(1) |^2$ is
proportional to the {\it deviation} of $\rho^2$ from the lower bound 
${3 \over 4}$.
Then, there is little room left for $\sum\limits_n 
|\tau^{(n)}_{1/2}(1) |^2$, as it has
been pointed out recently from a SR obtained for the subleading 
function $\xi_3(1)$
\cite{3r}~:

\beq
\label{7bise} \xi_3(1) = 2 \left [ \sum_n \Delta E_{3/2}^{(n)} 
|\tau_{3/2}^{(n)}(1)|^2 - \sum_n \Delta E_{1/2}^{(n)}
|\tau_{1/2}^{(n)}(1)|^2 \right ] \quad .
\eeq

\noi This sum rule, combined with Voloshin sum rule \cite{4bisr}

\beq
\label{8bise}
\overline{\Lambda} = 2 \sum_n \Delta E_{1/2}^{(n)} 
|\tau_{1/2}^{(n)}(1)|^2 + 4 \sum_n \Delta
E_{3/2}^{(n)}|\tau_{3/2}^{(n)}(1)|^2 \eeq

\noi yields
\bea
\label{9bise}
&& \sum_n \Delta E_{3/2}^{(n)}|\tau_{3/2}^{(n)}(1)|^2 = {1 \over 6} 
\left [ \overline{\Lambda} + \xi_3(1) \right ]
\\ && \sum_n \Delta E_{1/2}^{(n)}|\tau_{1/2}^{(n)}(1)|^2 = {1 \over 
6} \left [ \overline{\Lambda} - 2 \xi_3(1)
\right ] \quad . \label{10bise}
\eea

\noi Ignoring short distance QCD corrections, QCD sum rules predict, 
independently of all sum rule parameters
\cite{5bisr}

\beq
\label{11bise}
\xi_3(1) = {\overline{\Lambda} \over 3}
\eeq

\noi giving

\beq
\label{12bise}
{\sum\limits_n \Delta E_{1/2}^{(n)} |\tau_{1/2}^{(n)}(1)|^2 \over 
\sum\limits_n \Delta E_{3/2}^{(n)} |\tau_{3/2}^{(n)}(1)|^2} = {1
\over 4} \quad .  \eeq

\noi Since the L.S  coupling is small, we see that we have the same 
trend of inequality between $\sum\limits_n
|\tau_{3/2}^{(n)}(1)|^2$ and $\sum\limits_n|\tau_{1/2}^{(n)}(1)|^2$ 
as in equations (\ref{5e}) and (\ref{6e}).

\section{Uraltsev Sum Rule in Bakamjian-Thomas quark models.}
\hspace*{\parindent}
One of the aims of this note is to show that the SR (\ref{3e}) follows within
quark models \`a la Bakamjian and Thomas. Quark models of hadrons 
with a fixed number of
constituents, based on the Bakamjian-Thomas (BT) formalism 
\cite{4r,5r}, yield form
factors that are covariant and satisfy Isgur-Wise (IW) scaling 
\cite{6r} in the heavy
mass limit. In this class of models, the lower bound (\ref{4e}) was 
predicted some years
ago \cite{4r}. Moreover, this approach satisfies the Bjorken SR that relates
the slope of the IW function to the $P$-wave IW functions $\tau_{1/2}(w)$,
$\tau_{3/2}(w)$ at zero recoil \cite{7r}. In this approach were also 
computed the
$P$-wave meson wave functions and the corresponding inelastic IW 
functions \cite{8r}, and
a numerical study of $\rho^2$ in a wide class of models of the meson 
spectrum was
performed (each of them characterized by an Ansatz for the mass 
operator $M$, i.e. the
dynamics of the system at rest) \cite{9r}, together with a 
phenomenological study of the
elastic and inelastic IW functions and the corresponding rates for $B 
\to D, D^*,
D^{**}\ell \nu$. Moreover, the calculation of decay constants of 
heavy mesons within the
same approach was also performed \cite{10r}. \par

	The first demonstration of Uraltsev SR within the BT quark 
models is rather short, relying on formulas established in
ref. \cite{8r}. Two other demonstrations will follow that will 
exhibit the underlying physics. The starting point is
\cite{8r}~:

\beq
\label{7e}
		\tau^{(n)}_j(1)  = \int  {p^2dp \over (2\pi)^2}  \ \varphi^{(n)*}_j(p)  \ F_j(p)
\eeq

\noi where

\bea
\label{8e}
&&F_{1/2}(p) = - {1 \over 3\sqrt{3}} \left \{ \varphi (p) \ {p^2 
\over m+p_0} \left ( 3 +
{m \over p_0} \right ) + 2pp_0 \ {d\varphi \over dp} \right \} \nn \\
&&  F_{3/2}(p) = - {1 \over 3\sqrt{3}} \left \{ \varphi (p)
\ {p^2 \over m+p_0} \  {m \over p_0} + 2pp_0 \ {d\varphi \over dp} 
\right \}
\eea

\noi with the radial part of the $L=1$ wave functions normalized according to

\beq
\label{9e}
  		{1 \over 6\pi^2} \int  p^2dp \left [ p \varphi^{(n)}_j  (p) \right ]^2 = 1
\eeq

\noi and $m$, $p =| {\bf p} |$ and $p_0 = \sqrt{p^2+m^2}$ are the 
mass, momentum and
energy of the spectator quark. \par

 From (\ref{7e}), using closure in the sectors of definite $j = {1 
\over 2}$, ${3
\over 2}$ one finds (page 325 of ref. \cite{8r})~:

\beq
\label{12e}
		\sum_n  |\tau^{(n)}_j(1) |^2 = {3 \over 8\pi^2} \int  dp \ |F_j(p)|^2 \quad .
\eeq

\noi From (\ref{8e})-(\ref{12e}), the expression for the difference 
in the left-hand-side
of (\ref{3e}) can be integrated by parts, yielding, after some algebra~:

\beq
\label{13e}
  \sum_n  |\tau^{(n)}_{3/2}(1) |^2 - \sum_n  |\tau^{(n)}_{1/2}(1) |^2 
= {1 \over 8\pi^2}
\int  p^2dp \left [ \varphi (p) \right ]^2 = {1 \over 4}   \eeq

\noi where the last equality follows from the ground state wave 
function normalization
\cite{4r}. \par

Therefore, the SR (\ref{13e}) within the BT quark models provides a 
{\it rationale}  for
the lower bound $\rho^2  \geq  {3 \over 4}$  that was found within 
this class of models
\cite{4r}. The sum rule also establishes that the sum over the $j = 
3/2$ states dominates
over the one over the $j = 1/2$. \par

The second demonstration, that follows more closely Uraltsev proof, 
will illustrate quark-hadron duality. Let us first
remind the proof of Bjorken SR that was given in \cite{7r}. It was 
shown that the {\it spin averaged} hadronic tensor in
the BT formalism is, {\it in the heavy quark limit} for the active 
quark, identical to the free quark hadronic tensor~:

\begin{equation}
\label{12ebis}
\bar{h}_{\mu \nu}({\bf v}, {\bf v}') = \bar{h}_{\mu \nu}^{free \ 
quark}({\bf v}, {\bf
v}') \quad .
\end{equation}

\noi From this relation, Bjorken SR follows. In equation 
(\ref{12ebis}), the free quark tensor is

\beq
\label{13ebis}
\bar{h}_{\mu\nu}^{free\ quark}({\bf v}, {\bf v}') = {1 \over 2} 
\sum_{s_1,s'_1} \left [
\bar{u}_{s'_1}({\bf v}')\gamma_{\mu} u_{s_1}({\bf v}) \right ] \left 
[ \bar{u}_{s'_1}
({\bf v}') \gamma_{\nu} u_{s_1}({\bf v})\right ]^*\eeq

\noi and the hadronic tensor writes

\beq
\label{14tere}
\overline{h}_{\mu \nu}({\bf v}, {\bf v}') = {1 \over 2 J + 1} 
\sum_{\lambda} \sum_n <{\bf P}, \lambda |J_{\nu}|n, {\bf
P}'> <n,{\bf P}'|J_{\mu}|{\bf P}, \lambda >  \quad . \eeq

\noi where $J$, $\lambda$ are the spin and spin projection of the 
hadron of momentum ${\bf P}$.\par

  In BT models, the
hadronic tensor can be written \cite{7r}~:

\beq
\label{14ebis}
\bar{h}_{\mu\nu}({\bf v}, {\bf v}') = {1 \over 2J+1} \sum_{\lambda}
\sum_{s_{1f},s'_1,s_{1i}} \left [ \bar{u}_{s'_1}({\bf 
v}')\gamma_{\mu} u_{s_{1i}}({\bf
v}) \right ] \left [ \bar{u}_{s'_1} ({\bf v}') \gamma_{\nu} 
u_{s_{1f}}({\bf v})\right
]^* f_{s_{1f}s_{1i}}^{\lambda\lambda} \eeq

\noi where $f_{s_{1f}s_{1i}}^{\lambda\lambda}$ is the hadronic overlap~:
\beq
\label{15ebis}
f_{s_{1f}s_{1i}}^{\lambda\lambda} = \sum_{s_2} \int d^3{\bf p}_2 \
\psi_{s_{1f}s_2}^{\lambda*}\left ( {\bf P} - {\bf p}_2,{\bf p}_2 \right ) \
\psi_{s_{1i}s_2}^{\lambda} \left ( {\bf P} - {\bf p}_2, {\bf p}_2 
\right ) \quad . \eeq

\noi and (\ref{12ebis}) follows from (\ref{14ebis}) and 
(\ref{15ebis}). The wave function $\psi_{s_1,s_2}^{\lambda}({\bf 
P}-{\bf p}_2, {\bf
p}_2)$ is the internal moving ground state wave function, with the 
active quark labelled 1 and $\lambda$ being the spin
projection along some axis. It is defined by deleting the momentum 
conserving $\delta$-function from the total wave
function. In the BT model, it is obtained from a $P$-depending 
transformation on the {\it rest} internal wave function.
\par

To proceed like Uraltsev, one must generalize the hadronic tensor, allowing for
different velocities and angular momentum projections. Let us 
consider the {\it polarized}
hadronic tensor~:

\beq
\label{16ebis}
h_{\mu \nu}^{\lambda_i\lambda_f} ({\bf v}_i, {\bf v}_f, {\bf v}') = 
\sum_n <{\bf
P}_f,\lambda_f |J_{\nu}|n, {\bf P}' > < n, {\bf P}'|J_{\mu}|{\bf 
P}_i, \lambda_i >
\quad . \eeq

\noi In the BT formalism, this tensor writes, using closure and heavy 
mass limit \cite{7r}\footnote{The states $|n,
{\bf P}'>$ form a complete set of eigenfunctions at fixed ${\bf 
P}'$~: $\sum\limits_n |n, {\bf P}'><n,{\bf P}'| = 1$.}~:

\beq
\label{17ebis}
h_{\mu \nu}^{\lambda_i\lambda_f}  ({\bf v}_i, {\bf v}_f, {\bf v}') =
\sum_{s_{1f},s'_1,s_{1i}} \left [ \bar{u}_{s'_1} ({\bf v}') 
\gamma_{\mu}u_{s_{1i}} ({\bf
v}_i) \right ] \left [ \bar{u}_{s'_1} ({\bf v}') \gamma_{\nu}u_{s_{1f}} ({\bf
v}_f) \right ]^* f_{s_{1f}s_{1i}}^{\lambda_f\lambda_i}({\bf P}_i, 
{\bf P}_f)\eeq

\noi with the hadronic overlap

\beq
\label{18ebis}
f_{s_{1f}s_{1i}}^{\lambda_f\lambda_i}({\bf P}_i, {\bf P}_f) = 
\sum_{s_2} \int d^3 {\bf
p}_2 \ \psi_{s_{1f}s_2}^{\lambda_f*}\left ( {\bf P}_f - {\bf p}_2, 
{\bf p}_2 \right )
\  \psi_{s_{1i}s_2}^{\lambda_i}\left ( {\bf P}_i - {\bf p}_2, {\bf 
p}_2 \right )
\quad . \eeq

\noi In this expression $\psi_{s_{1i}s_2}^{\lambda_i} ( {\bf P}_i - 
{\bf p}_2, {\bf
p}_2 )$ ($i \to f$ likewise) is the internal moving ground state 
meson wave function,
and the active quark is labelled 1. \par

Let us choose, like Uraltsev, the vector meson $B^*$ as initial and 
final state, with
${\bf P}_i = 0$, $\lambda_i = 0$, $\lambda_f =  + 1$, and the vector 
current with $\mu
= \nu = 0$. We are thus considering the object

\beq
\label{19ebis}
h_{00}^{0,+1}({\bf v}_i, {\bf v}_f, {\bf v}') = 
\sum_{s_{1f},s'_1,s_{1i}}  \left [
\bar{u}_{s'_1}({\bf v}') \gamma_0u_{s_{1i}}(0) \right ]  \left [
\bar{u}_{s'_1}({\bf v}') \gamma_0u_{s_{1f}}({\bf v}_f) \right ]^*
f_{s_{1f}s_{1i}}^{+1,0}(0, {\bf P}_f)  \eeq

\noi to first order in ${\bf v}'$, ${\bf v}_f$. There are, in 
principle, two kinds of
terms contributing to this quantity~:

1) Spin-flip term coming from the active quark, i.e., from the quark 
current matrix
element at the desired order $\bar{u}_{s'_1}({\bf v}') \gamma_0 
u_{s_{1f}}({\bf v}_f)
\sim {\bf v}' \times {\bf v}_f$ while $\bar{u}_{s'_1}({\bf v}') \gamma_0
u_{s_{1i}}(0)$ cannot give a spin flip because ${\bf v}_i = 0$. At the
desired order, one can also take the hadronic overlap at ${\bf P}_i = 
{\bf P}_f = 0$~:

\beq
\label{20ebis}
h_{00}^{0,+1}(0, {\bf v}_f, {\bf v}') = \left
[ \bar{u}_{-1/2}({\bf v}') \gamma_0u_{-1/2}(0) \right ]  \left [
\bar{u}_{-1/2}({\bf v}') \gamma_0u_{+1/2}({\bf v}_f) \right ]^*
f_{s_{1f}s_{1i}}^{+1,0}(0, 0) \quad .
\eeq

\noi One obtains

\beq
\label{21ebis}
h_{00}^{0,+1}(0, {\bf v}_f, {\bf v}') = {1 \over 4 \sqrt{2}} \left ( 
< \downarrow
|i\sigma_1\cdot ({\bf v}' \times {\bf v}_f ) |\uparrow > \right )^* \eeq

\noi where the factor $1/\sqrt{2}$ comes from the hadronic overlap, 
and 1 labels the active quark. \par

2) Terms without spin-flip of the active quark. Then, to have a contribution to
(\ref{19ebis}), one needs to appeal to a Wigner rotation of the 
spectator quark 2,
giving a contribution $\sim {\bf p}_2 \times {\bf P}_f$. But, by 
integration, this term is zero, because there is no
other hadron momentum than  ${\bf P}_f$ -- in the hadronic overlap
there is no dependence on ${\bf P}'$. \\

We are then left with expression (\ref{21ebis}), that means that we 
have exact duality,
just like in the unpolarized, ${\bf P}_i = {\bf P}_f$ case~:

\beq
\label{22ebis}
h_{00}^{0,+1}(0, {\bf v}_f, {\bf v}') = \left [ h_{00}^{0,+1}(0, {\bf v}_f,
{\bf v}') \right ]_{free\ quark} \quad . \eeq

  We need now to compute the same hadronic tensor (\ref{16ebis}) in terms of the
phenomenological Isgur-Wise functions $\tau_j(w)$, within the same 
approximations. After
a good deal of algebra, we find, using the definitions of \cite{12r}, 
and taking into account that the states are not
normalized according to the usual normalization $< {\bf v}'|{\bf v} > 
= \sqrt{4v^0v'^{\ 0}} \  \delta
({\bf v}-{\bf v}')$ but by $< {\bf v}'|{\bf v} > = \delta ({\bf v}-{\bf v}')$,

\bea
\label{19e}
&&h_{00}^{0, +1}\left ( {\bf v}_f,{\bf v}',{\bf v}_i \right )  \cong 
v^z_f \ {1\over
\sqrt{2}} \ \left ( v'^{\, x}-iv'^{\, y} \right ) \\
&&\left [ C \left (0^+,j={\scriptstyle {1 \over 2}}\right )  +
C\left (1^+,j={\scriptstyle{1 \over 2}} \right )  + C\left (1^+,j= 
{\scriptstyle{3 \over 2}} \right )  + C\left
(2^+,j={\scriptstyle {3 \over 2}} \right ) \right ]  \nn
\eea
 
\noi where the different contributions are (a sum over a radial 
quantum number is implicit)

\bea
\label{20e}
&C\left ( 0^+,j={1 \over 2} \right )  = 0\qquad\qquad 
	&\qquad C\left ( 1^+,j={\scriptstyle{1 \over
2}}\right )  = - |\tau_{1/2}(1)|^2 \nn \\
&C \left ( 1^+,j={3 \over 2} \right ) = - \displaystyle{{1 \over 2}} 
|\tau_{3/2}(1)|^2	&\qquad C\left
(1^+,j={\scriptstyle {3 \over 2}}\right )  = {3 \over 2} 
|\tau_{3/2}(1)|^2		 \eea

\noi and the ground state does not contribute. One obtains,

\beq
\label{21e}
h_{00}^{0,+1}\left ( {\bf v}_f,{\bf v}',{\bf v}_i \right )  \cong 
v^z_f \ {1 \over
\sqrt{2}} \left ( v'^{\, x}-iv'^{\, y} \right )
\left [ \sum_n  |\tau^{(n)}_{3/2}(1) |^2 - \sum_n
|\tau^{(n)}_{1/2}(1)|^2 \right ] 	\ .  \eeq

\noi Identifying the expressions (\ref{21ebis}) and (\ref{21e}), 
Uraltsev SR follows. \\

Some words of caution about the general scope and limitations of 
Bakamjian-Thomas quark models are in order here. Both
zero order moment sum rules, the ones of Bjorken \cite{7r} and 
Uraltsev are satisfied by this class of models. However,
higher moment sum rules as Voloshin sum rule \cite{4bisr} are not 
satisfied. These higher moments sum rules seem to be
specific to the gauge nature of QCD. Anyhow, one limitation of BT 
models is the following, as exposed in \cite{4r}. The
Bakamjian-Thomas scheme was formulated to describe relativistic bound 
states with a fixed number of constituents, that
form representations of the Poincar\'e group. However, when one 
considers matrix elements of currents with one active
quark (the simplest Ansatz), these matrix elements are not covariant 
in general, although a main result of the formalism
is that they are covariant in the heavy quark limit. In the fact, one 
does not obtain a covariant expression for the
Voloshin sum $4 \sum\limits_n \Delta 
E_{3/2}^{(n)}|\tau_{3/2}^{(n)}(1)|^2 + 2 \sum\limits_n \Delta 
E_{1/2}^{(n)}
|\tau_{1/2}^{(n)}(1)|^2$, reflecting the non-covariance outside the 
heavy quark limit, by contrast to the Bjorken and
Uraltsev ones, that are covariant.

\section{The role of spectator quark Wigner rotations.}
 
Within quark models \`a la BT, the difference between $\tau^{(0)}_{3/2}(1)$ and
$\tau^{(0)}_{1/2}(1)$ follows from formulas (\ref{7e}), (\ref{8e}) 
(with a suitable phase convention,
$\tau_{3/2}^{(0)}$, $\tau_{1/2}^{(0)} \geq 0$)

\beq
\label{10e}
	\tau^{(0)}_{3/2}(1)  - \tau^{(0)}_{1/2}(1)  \cong {1 \over 
(2\pi)^2\sqrt{3}} \int
p^2dp \left [p \varphi^{(0)}_{L=1}(p) \right ]^* {p \over p_0+m} \ 
\varphi (p) \eeq

\noi where $\varphi_{1/2}(p) \cong \varphi_{3/2}(p) = 
\varphi^{(0)}_{L=1}(p)$ (assuming
small LS coupling) are the internal hadron wave functions at rest. We 
assume, as it is natural, that for the ground
state $\varphi_{L=1}^{(0)}(p)$ is positive. One finds that 
$\tau^{(0)}_{3/2}(1)$ is larger than $\tau^{(0)}_{1/2}(1)$
even in the limit of vanishing LS coupling. The difference 
(\ref{10e}) has a simple physical interpretation,
outlined in ref. \cite{9r}~: it is essentially due to the 
relativistic structure of the
matrix elements in terms of the wave functions. More precisely, it is 
due to the light
spectator {\it quark Wigner rotations}, i.e. a relativistic effect due to the
center-of-mass boost, and not due to the difference coming from the 
spin-orbit force
between the 1/2 and 3/2 internal wave functions at rest, which is 
small and has a rather
moderate effect. On the contrary, the difference (\ref{10e}) is quite 
large, at least
for the lowest $L=1$ states, since for a constituent quark mass $m 
\cong 0.3$~GeV, the
quantity ${p \over p_0+m}$ is of $O(1)$. \par

Expression (\ref{10e}), that comes from a specific relativistic 
effect, is to be
contrasted with the equality for any non-relativistic quark model 
with spin-orbit
independent potential \cite{11r}, also used in ref. \cite{12r}, that 
analizes $1/m_Q$
corrections~:

\beq
\label{11e}
				\tau^{(n)}_{3/2}(1)  = \tau^{(n)}_{1/2}(1) \quad .
\eeq
\vskip 3 truemm

Let us see how, in terms of internal wave functions {\it at rest}, the
Wigner rotation of the {\it spectator quark} is responsible for the 
difference between $\tau_{3/2}^{(n)}$ and
$\tau_{1/2}^{(n)}$ and for the non-vanishing of the r.h.s. of 
Uraltsev SR (\ref{13e}) within the BT formalism.
In the previous demonstration of Uraltsev SR, the Wigner rotations 
were hidden in the moving internal wave
functions, which themselves disappeared using completeness relations. 
We will now make those explicit by using the
internal wave functions at rest, that gives a feeling of how the 
difference $|\tau_{3/2}^{(n)}|^2 - |\tau_{1/2}^{(n)}|^2$
comes out in the l.h.s. of Uraltsev SR. Consider a meson with the 
active heavy quark labelled 1 and the spectator quark
labelled 2. In terms of internal wave functions, the current matrix 
element in the BT formalism writes (formula (27) of
ref. \cite{7r})~:

\bea
\label{14e}
&&	< {\bf v}'|V_{\mu}(0)|{\bf v} > \ = \sum_{s'_1s_1}  \bar{u}_{s'_1}
\gamma_{\mu}u_{s_1}   \int  d{\bf p}_2 \ {\sqrt{(p_i\cdot 
v)(p'_i\cdot v')} \over p^0_2}
\nn \\
&&\sum_{s'_2s_2}  \varphi'^*_{s'_1s'_2}({\bf k}'_2) \left [ D\left ( 
R'^{-1}_2   R_2
\right ) \right ]_{s'_2 s_2} \varphi_{s_1s_2}({\bf k}_2)	\quad .
\eea

\noi In this expression we see the basic ingredients of the model. 
There is a change of
variables of the quark momenta e.g. for the initial state $({\bf 
p}_1,{\bf p}_2) \to
({\bf P},{\bf k}_2)$, where {\bf P} is the center-of mass momentum, 
and ${\bf k}_2$ the
internal relative momentum, and likewise for the final state $({\bf 
p}'_1 ,{\bf p}'_2 )
\to ({\bf P}',{\bf k'}_2)$. The first term under the integral comes 
from the Jacobian of
this change of variables. The matrix element $u_{s'_1} \gamma_{\mu} 
u_{s_1}$ expresses
the fact that the quark 1 is the active heavy quark. The relation 
between e.g. $k_2$ and
$p_2$ is given by the boost $k^0_2  = v^0p^0_2  - v^zp^z_2$, $k^z_2 
= v^0p^z_2  -
v^zp^0_2$, $k^{x,y}_2  = p^{x,y}_2$, $v$ being the four-velocity of 
the initial state.
The wave functions $\varphi$ and $\varphi '$ are the initial and 
final internal wave
functions at rest, dependent only on the relative momenta and Pauli 
spinors. Finally,
the matrix $D(R'^{-1}_2 R_2)$ is the Wigner rotation acting on the 
spin of the spectator
quark 2 due to the product of the boosts on the initial and final 
states. Formula (\ref{14e}) leads to the difference
(\ref{10e}) and to the r.h.s. of Uraltsev SR (\ref{13e}). Expanding 
the fourth component vector current matrix element
between the ground state and $L=1$ states up to the first power of 
${\bf v}$, ${\bf v}'$ gives, from (\ref{14e}) (formula
(29) of ref. \cite{7r})~:

\beq
\label{15e}
< n({\bf v}')|V_0(0)|0({\bf v}) > \cong   {1 \over 2} ({\bf v}'- {\bf v})
\cdot \left  ( n \left | -i \left ( p^0_2  {\bf r}_2 + {\bf r}_2 
p^0_2 \right ) +
{i\left ( \sigma_2 \times {\bf p}_2 \right ) \over p^0_2+m} \right | 0 \right )
  \eeq

\noi where $|0({\bf v}) >$ stands for the ground state wave function 
in motion and
likewise $|0)$ for the internal ground state at rest in terms of 
Pauli spinors. The first
operator $-i(p^0_2 {\bf r}_2 + {\bf r}_2 p^0_2 )$, where ${\bf r}_2$ is the
operator $i {\partial \over \partial {\bf p}_2}$, comes from the 
variation of the
Jacobian factor and the variation of the argument ${\bf k}$ of the 
wave function, while
the second operator $\displaystyle{{i (\sigma_2 \times {\bf p}_2) 
\over p_2^0 + m}}$
is the Wigner rotation. Equation (\ref{15e}) becomes, in the 
non-relativistic limit, the matrix element of the
electric dipole operator, and leads to the difference (\ref{10e}) 
through the latter spin-dependent term. To demonstrate
Uraltsev SR, we are interested in the hadronic tensor

\beq
\label{16e}
h_{00}^{+10}({\bf v}_f,{\bf v}',{\bf v}_i)  = \sum_n  < B^{*(+1)}({\bf
v}_f)|V_0(0)|n({\bf v}') >< n({\bf v}')|V_0(0)|B^{*(0)}({\bf v}_i) >
\eeq

The ground state does not contribute to the sum rule over 
intermediate states in (\ref{16e}), in HQET and likewise in BT
quark models, that satisfy HQET. We have indeed demonstrated in ref. 
\cite{4r} (formulas (\ref{19ebis})-(\ref{22ebis}))
that BT quark models in the heavy quark limit satisfy HQET relations 
for all ground state form factors. More specifically,
in BT quark models, as follows after some algebra from (\ref{14e}), 
the contributions of the active quark
(\ref{21ebis}) cancels with the one of the spectator quark for the 
ground state. We are then left with the $L=1$
intermediate states for which we apply formula (\ref{15e}). \par

Defining the frame $v_i = (1,0,0,0)$, $v_f = (v^0_f
,0,0,v^z_f )$, the hadronic tensor can then be written, at first
order in the velocities ${\bf v}_f$ and ${\bf v}'$,

\beq
\label{17e}
h_{00}^{+10}({\bf v}_f,{\bf v}',{\bf v}_i) \cong {1 \over 4} \left ( 
B^{*(+1)} \left |
\left \{- v^z_f \left [ -i \left ( p^0_2 z_2+z_2 p^0_2 \right ) +
{i\left ( \sigma_2 \times {\bf p}_2 \right )_z \over p^0_2+m} \right 
] \right \}^+
\right | n \right )\eeq
$$\left ( n \left | \left \{ v'^{\, x} \left [ -i\left ( p^0_2 
x_2+x_2 p^0_2 \right ) +
{i\left ( \sigma_2 \times {\bf p}_2\right )_x \over p^0_2+m} \right 
] + v'^{\, y}\left
[-i\left (p^0_2  y_2+y_2 p^0_2 \right ) + {i\left ( \sigma_2 \times {\bf p}_2
\right )_y \over p^0_2+m} \right ] \right \} \right | B^{*(0)} \right )$$
\vskip 5 truemm

\noi where the $|n)$ states are $L=1$. The spin flip $B^{*(0)} \to
B^{*(+1)}$ can occur because of the Wigner rotation on the spectator 
light quark.  Using completeness  $\sum\limits_n
|n)(n| = 1$, two kinds of terms contribute~: crossed terms between a 
Wigner rotation and a
spin-independent operator, and products of two Wigner rotations. 
After some algebra, the
final result reads~:

\beq
\label{18e}
h_{00}^{+10}\left ( {\bf v}_f,{\bf v}',{\bf v}_i \right )  \cong  {1 
\over 4}  \ v^z_f
\ {1 \over \sqrt{2}} \left ( v'^{\, x}-iv'^{\, y} \right ) \quad . \eeq

\noi Making explicit the states $|n)$, equation (\ref{17e}) shows 
that the $L = 1$ states contribute to the left-hand
side of Uraltsev sum rule (eq. (\ref{21e})), since the operators in 
brackets are $\Delta L = 1$.\par

It may seem surprising that only a spectator quark operator appears 
in eq. (\ref{17e}), giving the same result as the
previous calculation (\ref{21ebis}), where only the active quark 
appeared. This is due to the fact that the right-hand
side of eqn. (\ref{21ebis}) or (\ref{18e}) comes out from a 
combination of three terms~: $S_1 + S_2 + P_2$, where $S(P)$
means the $S$-wave ($P$-wave) contribution and 1(2) the active 
(spectator) quark. It turns out that $S_1 = -
S_2 = P_2$, showing that one gets the same r.h.s. of the SR within 
both formalisms. The first demonstration underlines
duality, since the hadronic tensor is identical to the active quark 
tensor. The second demonstration underlines the
physical interpretation of the SR through the Wigner rotations, since 
the crossed terms $\Delta L = 1$, $\Delta S = 1$
in (\ref{17e}) provide the l.h.s. of the SR, giving the difference 
between $j = 3/2$ and $j = 1/2$.

\section{Phenomenological remarks.}
\hspace*{\parindent}
 From the calculations of ref. \cite{9r} in the BT formalism for a wide class of
potentials, one can see from Table 1 that Uraltsev SR converges 
rapidly, as well as
Bjorken's one, and are almost saturated by the $n = 0$ 
states\footnote{This fast convergence of the sum rules has also
been observed in QCD$_2$ in the $N_c \to \infty$ limit 
\cite{16bisr}.}. \par \vskip 5 truemm

\noindent  \begin{tabular}{|c|c|c|c|}
\hline
Quark-antiquark 	&Godfrey, Isgur \protect{\cite{13r}} &Cea, 
Colangelo, &Isgur, Scora, \\
Potential &($Q\bar{Q}$, $Q\bar{q}$, $q\bar{q}$)	&Cosmai, Nardulli 
\protect{\cite{14r}}
&Grinstein, Wise \protect{\cite{15r}} \\
\hline
$|\tau^{(0)}_{1/2}(1) |^2$	&0.051	&0.004	&0.117 \\
\hline
$|\tau^{(0)}_{3/2}(1) |^2$	&0.291	&0.265	&0.305 \\
\hline
${1 \over 4}  +|\tau^{(0)}_{1/2}(1) |^2$ +   &0.882	&0.790
&1.068 \\
$2|\tau^{(0)}_{3/2}(1) |^2$ & & & \\
\hline
$ \rho^2$   	&1.023	&0.98$\ $	&1.283 \\
\hline
$|\tau^{(0)}_{3/2}(1) |^2$   &0.240	&0.261	&0.233 \\
$- |\tau^{(0)}_{1/2}(1) |^2$ & & & \\
\hline
   \end{tabular}
\par \vskip 3 truemm

\noindent {\bf Table 1 : } Contribution of the lowest $L=1$ states to 
the Bjorken and Uraltsev Sum Rules
and the slope of elastic IW function in BT quark models for different
potentials. \par \vskip 5 truemm

\baselineskip=20 pt
The Godfrey and Isgur potential \cite{13r} is the one that describes 
the meson spectrum
in the most complete way, from light meson spectroscopy to heavy quarkonia. The
agreement of the contribution of lowest $n = 0$ states with the 
right-hand-side of the SR
(\ref{13e}) is quite striking. Within the BT class of quark models, 
one gets \cite{9r} a
value $\rho^2 \cong 1$, not inconsistent with present experimental 
data on the $\xi (w)$
slope, and also, consistently, with small values for 
$\tau^{(n)}_{1/2}(1)$. \par

It is interesting to remark that, among the three potential models 
quoted in Table 1, only the more complete one by
Godfrey and Isgur contains a L.S coupling. There are indeed in this 
case L.S splittings ($M_{3/2}^{(n)}$ different
from $M_{1/2}^{(n)}$), and the wave functions are perturbed also by 
this piece of the interaction, giving a different
behavior for the wave functions $\varphi_{3/2}^{(n)}(p)$ and 
$\varphi_{1/2}^{(n)}(p)$. The other models have neglected the
L.S splitting, although, due to the Wigner rotations, 
$\tau_{3/2}^{(n)}(w)$ is, of course, different from
$\tau_{1/2}^{(n)}(w)$ even for these latter potentials. However, even 
in the case of the Godfrey-Isgur potential, the L.S
force is small.

  In Table 2 we compare the predictions of the BT quark models for the different
semileptonic decays. While the BR for the modes $B \to D_2\ell \nu$ 
and $B \to D_1({3
\over 2} )\ell \nu$ have the right order of magnitude, and are 
consistent with experiment
within $1\sigma$, the trend of the ratio $D_1\left ( {3 \over 
2}\right )/D_2\left ( {3 \over 2}\right )$ is opposite to
experiment. This moderate disagreement could be explained by $1/m_Q$ 
corrections \cite{17r}. However,
in the case of the $j = {1 \over 2}$ the disagreement is very strong. 
QCD in the heavy quark limit predicts,
according to Uraltsev SR, that the $j = {3 \over 2}$ states are 
dominant over the $j = {1 \over 2}$.
This general trend could be hardly reversed by the small hard QCD
corrections to Uraltsev \cite {1r} and Bjorken \cite{17r} sum rules. 
As to the $1/m_Q$ corrections \cite{12r}, their
magnitude is poorly known, since the numerical estimate of ref. 
\cite{12r}, although the formalism is completely general,
relies on a large number of dynamical hypotheses. \par \vskip 5 truemm

\noindent
   \begin{tabular}{|c|c|c|c|c|}
\hline
Quark-antiquark 	&Godfrey-	&Cea et al.	&Isgur et al. 
	&Expt. \\
Potential&Isgur & & &\\
\hline
$B \to D\ell \nu$	&2.36 \%	&2.45 \%	&1.94 \% 
	&$(2.1 \pm 0.2) \%$ \\
\hline
$B \to  D^*\ell \nu$	&6.86 \%	&7.02 \%	&6.07 \% 
	&$(5.3 \pm 0.8) \%$ \\
\hline
& & & &(a) $(2.4 \pm 1.1) \times 10^{-3}$ \\
$B \to D_2\left ( {3 \over 2} \right ) \ell \nu$	&$7.0 \times 
10^{-3}$ 	&$6.5 \times 10^{-3}$	&$7.7 \times 10^{-3}$
&(b) $(4.4 \pm 2.4) \times 10^{-3}$ \\
& & & &(c) $(3.0 \pm 3.4) \times 10^{-3}$ \\
\hline
& & & &(a) $(7.0 \pm 1.6) \times 10^{-3}$ \\
$B \to D_1\left ( {3 \over 2} \right  ) \ell \nu$	&$4.5 \times 
10^{-3}$	&$4.2 \times
10^{-3}$ &$4.9 \times 10^{-3}$ &(b) $(6.7 \pm 2.1) \times 10^{-3}$ \\
& & & &(c) $(5.6 \pm 1.6 ) \times 10^{-3}$ \\
\hline
$B \to D_1\left ( {1 \over 2} \right  )	\ell \nu$ &$7 \times 10^{-4}$	&$4
\times 10^{-5}$ &$1.3 \times 10^{-3}$ &$(2.3 \pm 0.7) \times 10^{-2}$ \\
\cline{1-4}
$B \to D_0 \left ( {1 \over 2} \right  )	\ell \nu$	&$6
\times 10^{-4}$  &$4 \times 10^{-5}$ &$1.1 \times 10^{-3}$ &$\left [ 
D_0 \left ( {1
\over 2} \right  ) +D_1 \left ( {1 \over 2} \right  ) \right ]$ \\
\hline
\end{tabular}
\par \vskip 5 truemm

\noindent {\bf Table 2 : } Branching ratios in BT quark models for 
different potentials. The ex\-pe\-ri\-men\-tal BR for
$B \to D_2\left ({3 \over 2} \right )\ell \nu$ and $B \to D_1\left 
({3 \over 2} \right )\ell \nu$ come from ALEPH (a),
DELPHI (b) and CLEO (c) data \protect{\cite{16r}}, with the errors 
added in quadrature. The last
entry corresponds to DELPHI data for the wide states. \par \vskip 5 truemm

Another strong experimental indication of large branching ratios of a 
broad resonance $D_1\left ( {1 \over 2} \right )$
is the non-leptonic decay $B \to D_1^0\left ( {1 \over 2} \right ) 
\pi$ which is found larger than the $B \to D_J \left
( {3 \over 2}\right ) \pi$ \cite{18r}. Factorization is reasonable in 
such a mode and, consequently, once again, this
experimental result seems to contradict that $|\tau_{3/2}(1)| > 
|\tau_{1/2}(1)|$. \par

The serious problem for the decays $B \to D_{0,1}({1 \over 2}) \ell 
\nu$ goes beyond the specific BT quark models and
appears to be, more generally, a problem between experiment and the 
heavy quark limit of QCD.

\section{Conclusion.}
\hspace*{\parindent}
We have shown that the sum rule proved recently by Uraltsev in the 
heavy quark limit of QCD holds in
relativistic quark models \`a la Bakamjian and Thomas. Its physical 
interpretation is the
Wigner rotation of the spectator light quark spin, and not a possible 
LS perturbation.
We have underlined that, since $|\tau_{3/2}(1)| > |\tau_{1/2}(1)|$ 
\cite{19r}, there is a serious
problem between theory and experiment for the decays $B \to 
D^*_{0,1}(broad) \ell \nu$. This problem goes beyond the BT
quark models and appears to be a general one, within the heavy quark 
limit of QCD. \\

\noi {\large \bf Acknowledgements.} \\

We acknowledge very interesting discussions with Nikolai Uraltsev and 
also his careful reading of the manuscript. We
acknowledge partial support from TMR-EC Contract No. ERBFMRX-CT980169.

\end{document}